\documentclass[a4paper,12 pt]{article}
 \usepackage{times}
 \usepackage{amsmath, amssymb, amsthm}  
 \usepackage{graphicx} 
 \usepackage{float}
 \usepackage[margin= 0.75 in, letterpaper]{geometry} 

 \usepackage{setspace}  
 \usepackage{indentfirst}
 \usepackage{threeparttable}

 \usepackage{rotating}
 \usepackage{lscape} 
 \usepackage[natbibapa]{apacite} 
 \usepackage{multirow}
 \usepackage[labelsep=period]{caption}
\usepackage{hyperref}

\begin{document}  
\title{Bayesian compartmental modelling of MRSA transmission within hospitals in Edmonton, Canada}
\author{
	Ruoyu Li$^{1\ast}$, 
	Rob Deardon$^{2,3,1}$,
	Na Li$^{1,4,5}$,
	John Conly$^{6,7,8}$,
	Jenine Leal$^{1,6,8}$
	\and
        \small$^{1}$Department of Community Health Sciences, University of Calgary, Calgary, Canada.\and
	\small$^{2}$Department of Mathematics and Statistics, University of Calgary, Calgary, Canada.\and
	\small$^{3}$Faculty of Veterinary Medicine, University of Calgary, Calgary, Canada.\and
	\small$^{4}$Department of Computing and Software, McMaster University, Hamilton, Canada.\and 
	\small$^{5}$Centre for Health Informatics, Cumming School of Medicine, University of Calgary, Calgary, Canada.\and 
        \small$^{6}$Department of Microbiology, Immunology, and Infectious Diseases, University of Calgary, Calgary, Canada.\and
        \small$^{7}$Department of Medicine, University of Calgary and Alberta Health Services, Calgary, Canada.\and
        \small$^{8}$Infection Prevention and Control, Alberta Health Services, Calgary, Canada. \and
	\small$^\ast$Corresponding author. Email: ruoyu.li@ucalgary.ca \and
}
\date{}

\maketitle 

\section*{Abstract}
\noindent Methicillin-resistant $Staphylococcus$ $aureus$ (MRSA) is a bacterium that leads to severe infections in hospitalized patients. Previous epidemiological research has focused on MRSA transmission, but few studies have examined the influence of both hospital-acquired MRSA (HA-MRSA) and community-acquired MRSA (CA-MRSA) on MRSA spread in hospitals. In this study, we present a unique compartmental model for studying MRSA transmission patterns in hospitals in Edmonton, Alberta. The model consists of susceptible individuals, patients who have been colonized or infected with HA-MRSA or CA-MRSA, and isolated patients. We first use Bayesian inference with Markov chain Monte Carlo (MCMC) algorithms to estimate the posterior mean of parameters in the full model using data from hospitals in Edmonton. Then we develop multiple sub-models with varying assumptions about the origin of new MRSA colonization. We also estimate transmission rates in hospitals.

Keywords: Bayesian inference, colonization and infection, compartmental models, MRSA transmission 

\section{Introduction}
$Staphylococcus$ $aureus$ ($S. aureus$) is the most common Gram-positive bacterium in humans, with around 20\% carrying it persistently and 60\% carrying it intermittently \citep{nasal}. This bacterium can cause various infections, including surgical wound infections, bloodstream infections, and pneumonia, which can be treated with antibiotics like methicillin, oxacillin, flucloxacillin and dicloxacillin. However, with the increased use of antibiotics, $S. aureus$ strains have developed resistance, resulting in methicillin-resistant $S. aureus$ (MRSA) infections, which account for approximately 26\% of all $S. aureus$ cases and lead to severe morbidity and mortality among patients \citep{mrsarate, MRSA1}. In 2019, MRSA infections were responsible for more than 100,000 deaths and 3.5 million disability-adjusted life years worldwide \citep{death, ssti}. 

The prevalence of MRSA infections varies greatly among diverse populations in different regions and countries, ranging from less than 1\% to more than 50\% \citep{europe, africa}. In Canada, the first nosocomial MRSA outbreak occurred in 1978, and the first community-associated MRSA case was recorded in the early 2000s \citep{outbreak, sysreview}. In 1999, the mean MRSA prevalence rate (including colonized and infected) in acute care hospitals across Canada was 2.0 per 1,000 admissions \citep{meanrate}. MRSA prevalence rate has risen in both clinical and non-clinical settings in Alberta, Ontario, Quebec, and Saskatchewan \citep{alberta12, sysreview}. Between 2015 and 2019, healthcare-associated MRSA bloodstream infections increased by 15.4\% in Canada, from 0.39 to 0.45 cases per 10,000 patient-days, and community-associated MRSA bloodstream infections increased by 126.3\%, from 0.19 to 0.43 cases per 1,000 patient-admissions \citep{antimicrobialrate}.

Risk factors for MRSA colonization in hospitals include 1) severe underlying illness or comorbid conditions, 2) exposure to broad-spectrum antimicrobials, implanted foreign bodies, and invasive therapies like indwelling urinary catheters, 3) prolonged hospital stay, and 4) frequent contact with healthcare personnel or the healthcare system including long-term care facilities \citep{riskfactor}. In addition to these variables, non-intact skin, such as abrasions or incisions, might contribute to colonization by MRSA and subsequent MRSA infections. Intensive care unit patients are more prone to MRSA infections due to their weakened immune systems and frequent invasive procedures \citep{icu}. As a result, MRSA is regarded as a significant nosocomial pathogen, and patients infected while hospitalized are referred to as having hospital-acquired MRSA (HA-MRSA) infections. 

MRSA can spread in hospitals through direct contact, such as skin-to-skin contact, particularly in areas with infected wounds, and people with compromised immune systems are more vulnerable to MRSA colonization and infection. Contaminated medical equipment can also be a vector to propagate MRSA \citep{skin}. Furthermore, there is some transmission pressure during hospitalization because around 1.9\% of patients have been colonized with MRSA before their arrival, depending on local epidemiology \citep{transjapan}. Meanwhile, the frequency of MRSA in the community has increased significantly among non-hospitalized people or those who have not had a medical treatment within the last year, classified as community-acquired MRSA (CA-MRSA) infections. CA-MRSA spreads through person-to-person contact or contaminated fomites such as towels and clothes, and people living in congested environments, such as jails, have a higher risk of infection \citep{correctional}. 

Compartmental models have frequently been used to analyze MRSA transmission patterns in hospitals and in the community. They propose that populations are separated into compartments, each indicating a particular MRSA status, such as susceptible, colonized, infected, and recovered (or removed). An individual who is susceptible to MRSA will initially remain in the susceptible compartment of the study, if the individual later becomes colonized, that individual may migrate to the colonization compartment. Every patient who stays in the recovered (or removed) compartment demonstrates to others that the individual is no longer transferring the bacterium. In general, the sequence of the compartments corresponds to how patients move between them \citep{invasion, datadriven, stochaMRSA}.

In compartmental models, population-based assumptions are usually held, as we track groups in the population and determine the average values of patients in a given state. It is common to assume that the population has homogeneous mixing and that there is no variation among individuals, then every patient in the population has the same potential to spread MRSA \citep{introinfect}. Both deterministic and stochastic models can be used to study MRSA transmission. Deterministic models use a system of ordinary or partial differential equations to describe how the numbers or proportions of individuals in different states change over time. It is usually straightforward to obtain numerical solutions given a set of parameters, but the epidemic created by a deterministic model is not random. Stochastic models typically use a random probability distribution to describe both infection events and the latent and/or infectious periods of MRSA. They can be especially useful when populations are small, such as in an intensive care unit in a hospital. MRSA transmission cannot be completely captured by a model, hence stochastic models are preferred over deterministic models for capturing variations. The number of infected individuals in a stochastic model varies at random based on a distribution that accounts for the uncertainty and inherent randomness in the population and pathogen, such as differences in environmental or genetic factors that can affect epidemiological behaviour. So, there will be modest discrepancies in the estimated parameters between utilizing deterministic models and stochastic models, and these differences represent the uncertainty in the MRSA transmission \citetext{\citealp[Chapter 3]{mathepi}; \citealp[Chapter 6]{modelinfect}; \citealp{introinfect}; \citealp{simcompart}}. 

Some studies employed these models to study HA-MRSA dissemination. For example, \cite{nosoMRSA} developed both deterministic and stochastic compartmental models to analyze nosocomial MRSA infections. The models consisted of various compartments representing different classes of individuals, such as uncolonized patients, colonized patients, uncontaminated healthcare workers (HCWs), contaminated HCWs, and bacterial load in the environment. They used the deterministic model to estimate transmission parameters and used the stochastic model to take environmental contamination into account, showing that the change of colonized patients and contaminated HCWs in the stochastic model oscillate around the values in the deterministic model. Similarly, \cite{stochaMRSA} used a stochastic model to study MRSA transmission in a hospital ward, where patients were divided into various groups based on their MRSA status (susceptible, undetected MRSA colonizations, detected MRSA colonizations, undetected MRSA infections, and detected MRSA infections). The study estimated the transmission rate depending on the cumulative number of patients in those groups. Another study by \cite{antiMRSA} created both deterministic and stochastic models to investigate the effect of antibiotic exposure and environmental contamination on MRSA transmission in hospitals. The models included different compartments representing uncolonized patients with or without antibiotic exposure, colonized patients with or without antibiotic exposure, uncontaminated or contaminated HCWs, and free-living bacteria. The results from the deterministic model emphasized the importance of antibiotic stewardship. The stochastic model captured the change in MRSA infection in small populations and concluded that environmental cleaning played a crucial role in reducing the spread of MRSA. 

Some research used compartmental models to investigate CA-MRSA transmission. For instance, \cite{datadriven} developed an age-structured deterministic susceptible-exposed-infected-susceptible (SEIS) model to analyze skin and soft tissue infections caused by CA-MRSA in the community. The model consisted of susceptible individuals with no prior colonization with CA-MRSA, those colonized for the first time, those infected for the first time, and those with a history of past infections. \cite{uspopulationMRSA} employed a deterministic susceptible-colonized-infected-recovered-susceptible (SCIRS) compartmental modelling framework to investigate the spread of CA-MRSA among the US population. Moreover, \cite{invasion} used a deterministic compartmental model to analyze the spread of CA-MRSA in the hospital setting, and the model consisted of five compartments representing susceptible patients, patients colonized with CA-MRSA, patients infected with CA-MRSA, patients colonized with HA-MRSA, and patients infected with HA-MRSA. 

Here, we develop a population-based stochastic compartmental model to investigate MRSA spread across hospitals in Edmonton, Alberta, using data that include the aggregate number of patients colonized or infected with HA-MRSA or CA-MRSA per month to estimate MRSA transmission rate within hospitals. The contribution of this study is that our model gives a clear framework for how MRSA spreads among susceptible persons and patients colonized or infected with HA-MRSA and CA-MRSA, which is easily understandable to medical researchers, healthcare professionals, and the general public. The findings show that not only do patients colonized or infected with HA-MRSA in hospitals result in more susceptible individuals to become new patients colonized with HA-MRSA, but patients admitted to hospitals colonized or infected with CA-MRSA also increase the risk of MRSA transmission, which may provide policymakers with guidance in implementing targeted interventions to reduce MRSA transmission in hospitals. 

Section \ref{sec:epidata} gives information on data sources, and we primarily use data from hospitals in Edmonton to conduct this analysis. The provincial capital of Alberta is located in Edmonton, which has 14 hospitals and offers specialized referral services, such as organ transplant coverage for northern Alberta \citep{edmonton}. Section \ref{sec:model} describes the MRSA transmission model in hospitals, including its structure and behaviour. In Section \ref{sec:method}, we present the Bayesian inference and Markov chain Monte Carlo (MCMC) methods used in the study to estimate the posterior mean of parameters in the full MRSA transmission model. Bayesian inference incorporates the uncertainty in unobserved compartments or incomplete data into the parameter posterior distributions, increasing the model's flexibility \citep[Chapter 2]{predinter, bayesdata}. Then we fit multiple sub-models of the full model with various assumptions about how different patient groups influence MRSA transmission among susceptible individuals when they become new patients colonized with HA-MRSA, in order to determine which patient group has a greater impact on MRSA transmission. Section \ref{sec:result} shows results of the data analysis, focusing on both the full model and one sub-model. Section \ref{sec:discussion} covers the discussion and future work.

\section{Epidemiological data \label{sec:epidata}}
This study uses aggregate monthly data from Edmonton hospitals from the provincial surveillance data management system within Alberta Health Services's (AHS) Infection Prevention and Control (IPC) program \citep{provsurv}. They contain patient information related to Antimicrobial Resistant Organisms (ARO) in Alberta and record the incident number of MRSA infections or colonizations identified in hospitals. AHS and Covenant Health provide acute care services to over 4.5 million Albertans in 110 acute care and acute rehabilitation facilities, and the IPC surveillance program conducts ARO surveillance at all of these hospitals \citep{healthplan}.\\
\noindent \textbf{Patient population:} Individuals who have been admitted to AHS and Covenant Health acute and acute tertiary rehabilitation care facilities between September 2018 and September 2023. \\
\noindent \textbf{Case definition:} An initial case is a laboratory-confirmed MRSA from a body site that has been detected as positive for MRSA at the time of admission or during hospitalization.\\
\noindent \textbf{Inclusion criteria:} 1) MRSA cases identified for the first time while a patient was admitted to a facility under surveillance; 2) MRSA cases identified for the first time in the emergency department in patients who are subsequently admitted to a facility under surveillance (i.e. acute and acute tertiary rehabilitation facilities). \\
\noindent \textbf{Exclusion criteria:} 1) Patients with a previous incident MRSA record (unless identified with a different strain of MRSA); 2) MRSA positive cases identified by the laboratory among patients who were not admitted at the time of specimen collection or who were not subsequently admitted as an inpatient following their emergency department visit.

Cases are classified as hospital-acquired, healthcare-associated and community-acquired MRSA cases based on the time of identification and the patient's history of healthcare encounters. \\
\noindent \textbf{Hospital-acquired MRSA:} Newly-identified MRSA positive after the third calendar day of admission (first calendar day: day of admission) with the following criteria: 1) No previously identified MRSA colonization at the time of admission; 2) no MRSA positive cultures in the past; 3) no present or incubating MRSA infection on admission; 4) evidence shows the case is attributable to the facility under surveillance if identified within three calendar days. \\
\noindent \textbf{Healthcare-associated MRSA:} Newly-identified MRSA positive on the day of admission (first calendar day) or the day after admission (second calendar day) to an inpatient location with previous admissions or healthcare encounters in the last 12 months outside of the 14-day hospital-acquired attribution window.\\
\noindent \textbf{Community-acquired MRSA:} Newly-identified MRSA positive on the day of admission (first calendar day) or the day after admission (second calendar day) to an inpatient location without previous admissions or healthcare encounters in the last 12 months \citep{provsurv}.

\begin{figure}[htbp]
  \centering
  \includegraphics[width=16cm]{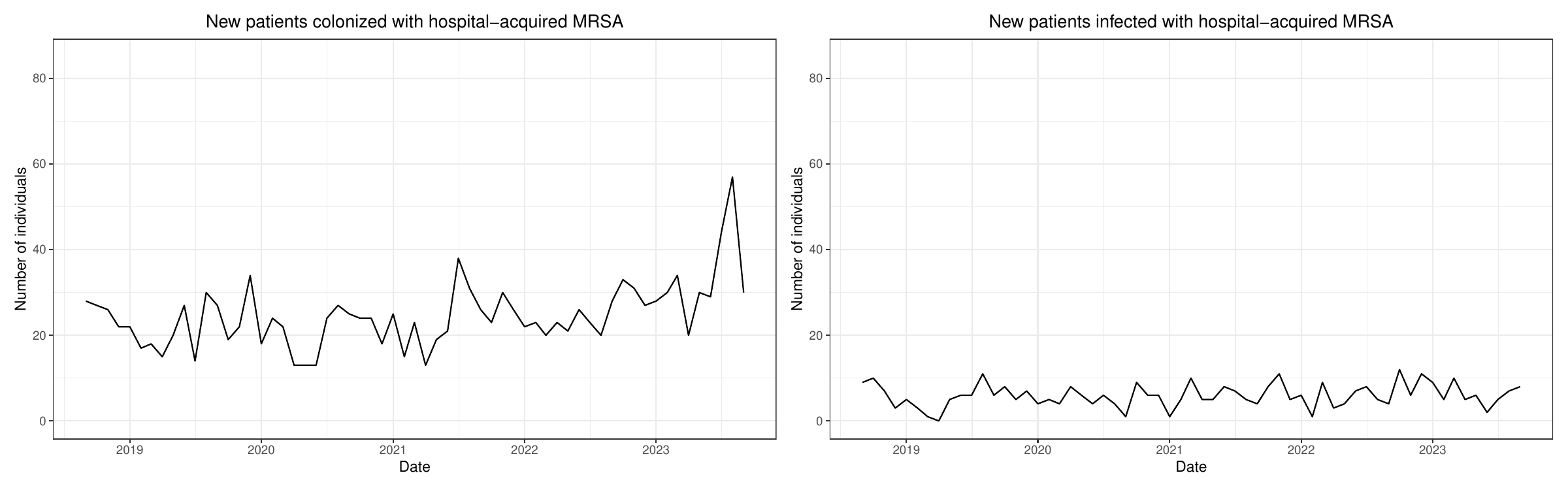}
  \caption{The number of patients in hospitals in Edmonton who were newly colonized with hospital-acquired MRSA (left) and infected with hospital-acquired MRSA (right) from September 2018 to September 2023.}
  \label{fig:ha-mrsa}
\end{figure}

\begin{figure}[htbp]
  \centering
  \includegraphics[width=16cm]{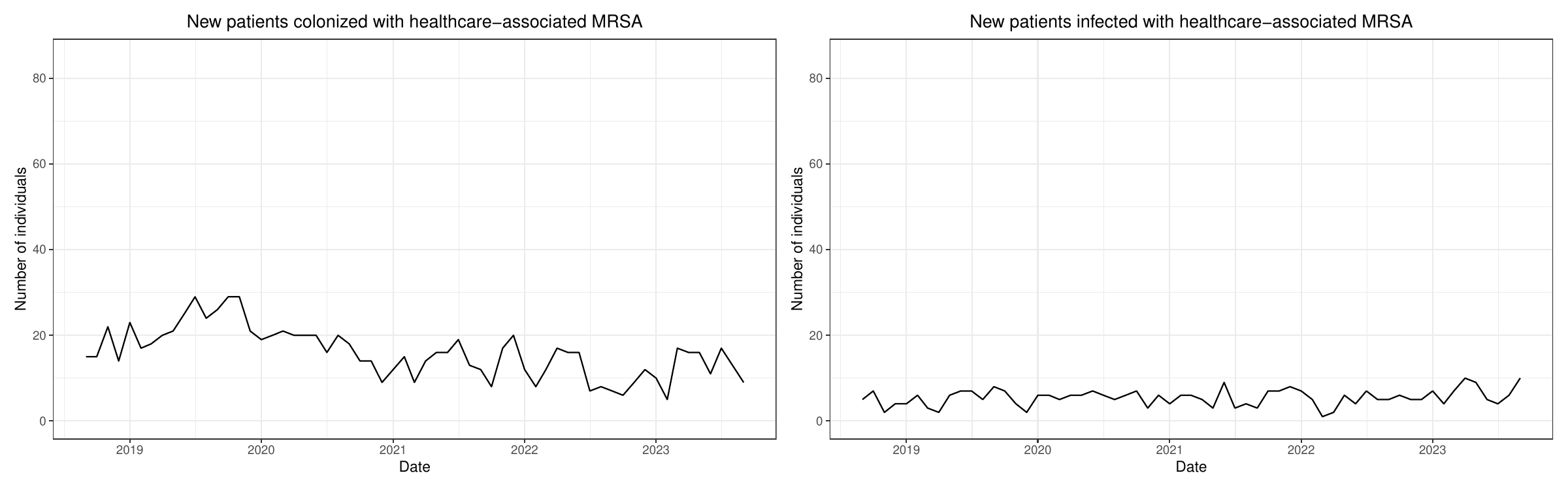}
  \caption{The number of patients in hospitals in Edmonton who were newly colonized with healthcare-associated MRSA (left) and infected with healthcare-associated MRSA (right) from September 2018 to September 2023.}
  \label{fig:hca-mrsa}
\end{figure}

\begin{figure}[htbp]
  \centering
  \includegraphics[width=16cm]{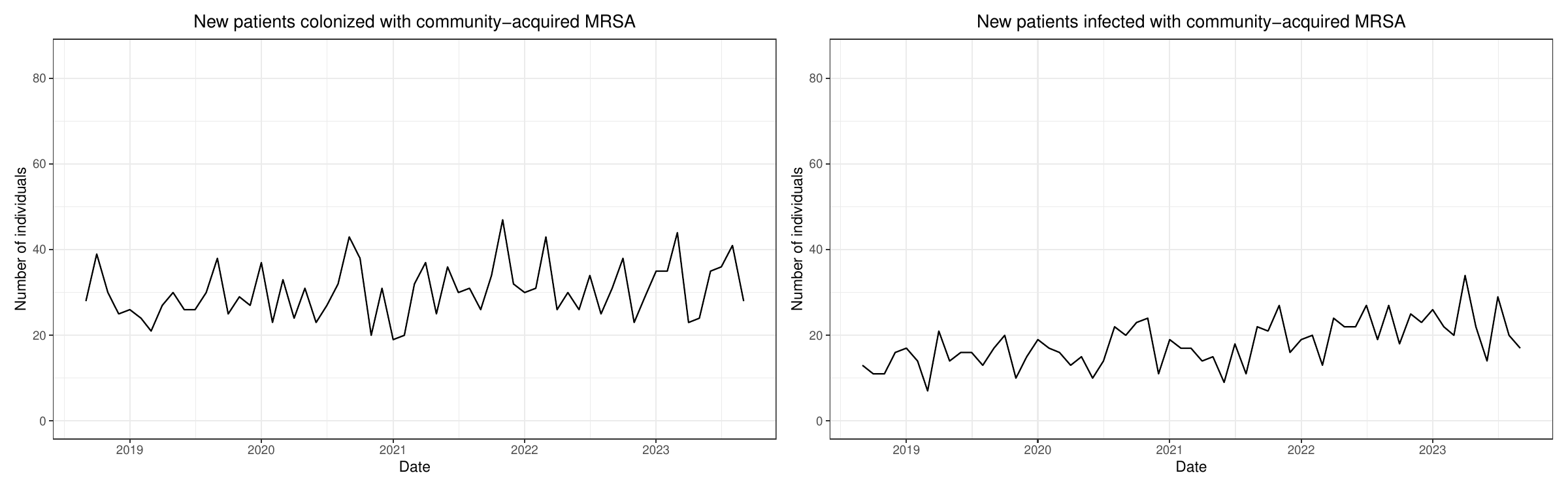} 
  \caption{The number of patients in hospitals in Edmonton who were newly colonized with community-acquired MRSA (left) and infected with community-acquired MRSA (right) from September 2018 to September 2023.}
  \label{fig:ca-mrsa}
\end{figure}

\begin{figure}[htbp]
  \centering
  \includegraphics[width=16cm]{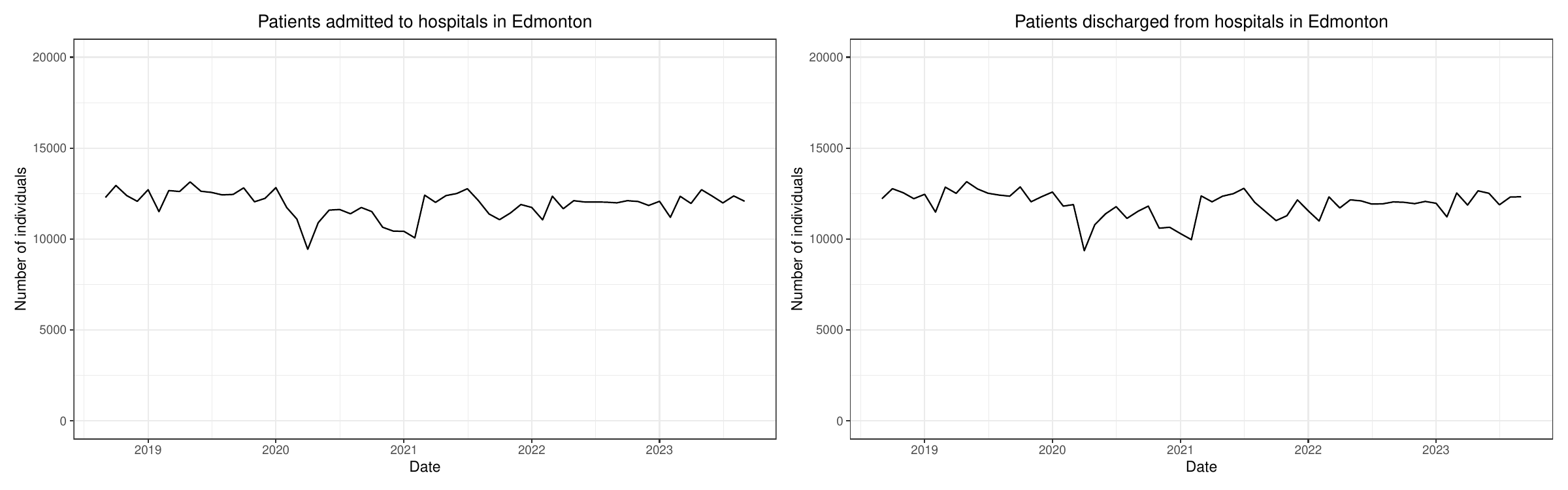}
  \caption{The number of patients admitted to (left) or discharged from (right) hospitals in Edmonton from September 2018 to September 2023.}
  \label{fig:admit}
\end{figure}

Figures \ref{fig:ha-mrsa} to \ref{fig:ca-mrsa} indicate the number of patients colonized or infected with hospital-acquired MRSA, healthcare-associated MRSA, or community-acquired MRSA per month during the research period, respectively. Between September 2018 and September 2023, there were about 4 times as many new hospital-acquired MRSA colonizations (mean = 24.6) as there were hospital-acquired MRSA infections (mean = 5.9), as shown in Figure \ref{fig:ha-mrsa}. The number of new patients colonized with hospital-acquired MRSA per month fluctuated before 2023, then increased substantially in August 2023 before declining, and there were also some modest changes in the number of new patients with hospital-acquired MRSA infections during the research period. 

According to Figure \ref{fig:hca-mrsa}, the number of patients colonized with healthcare-associated MRSA (mean = 15.9) is 3 times higher than the numberstocMCMC of patients infected with healthcare-associated MRSA (mean = 5.4). Additionally, the number of new patients colonized with healthcare-associated MRSA decreases from April 2019 to November 2019 and then continues to fluctuate for the remainder of the period. Simultaneously, the number of patients infected with healthcare-associated MRSA varies slightly. The number of patients colonized with community-acquired MRSA (mean = 30.3) in hospitals in Edmonton is 1.7 times higher than the number of patients infected with community-acquired MRSA (mean = 18.1). Additionally, we can see a greater fluctuation in patients colonized with community-acquired MRSA compared to patients infected with community-acquired MRSA during the study period (Figure \ref{fig:ca-mrsa}). 

In Figure \ref{fig:admit}, it appears that there is a considerable decline in April 2020 in the number of patients admitted to (mean = 11918) and discharged from (mean = 11920) hospitals. Both of them follow a similarly stationary pattern from September 2018 to September 2023. Overall, community-acquired MRSA colonizations outnumber those of hospital-acquired MRSA and healthcare-associated MRSA. We also use the Kruskal-Wallis test \citep{ranks} to detect seasonality and concluded that there are no seasonal changes in these patient groups ($p > 0.05$). Because both healthcare-associated MRSA and community-acquired MRSA cases are identified within two days of admission, and it is difficult to confirm that patients with healthcare-associated MRSA acquired the bacteria in hospitals, even if they have previously been admitted to a facility under surveillance during the previous 12 months, we combine them and classify them as community-acquired MRSA cases in the model. Our model will then include patients colonized or infected with hospital-acquired MRSA (HA-MRSA) and community-acquired MRSA (CA-MRSA).

\section{MRSA transmission model \label{sec:model}}
\begin{figure}[htbp]
  \centering 
  \includegraphics[width=10cm]{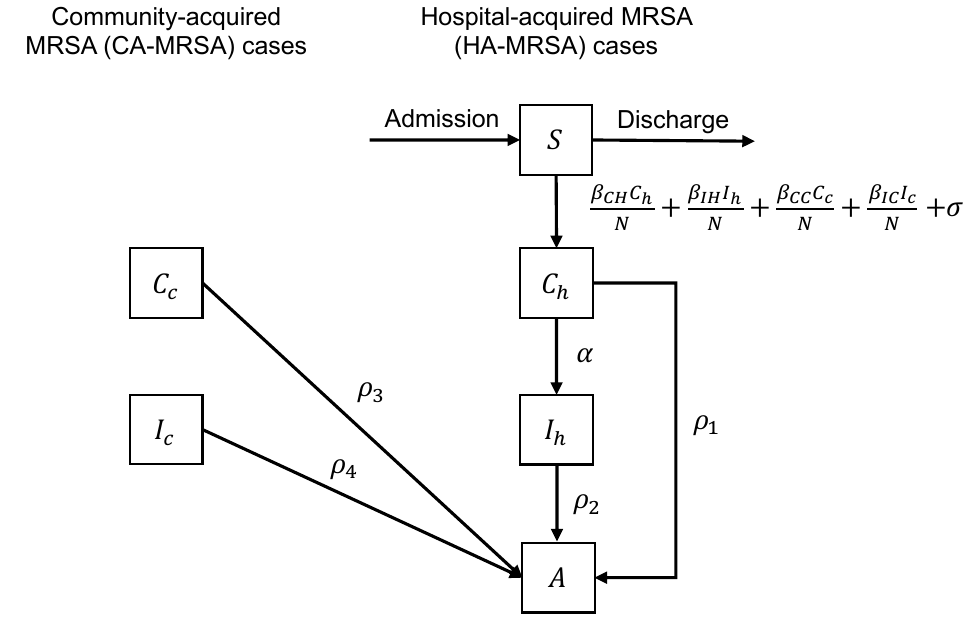}
  \caption{The full compartmental model of the transmission dynamics of MRSA in hospitals. In the model, $S$ is the number of susceptible individuals, $C_h$ is the number of patients colonized with HA-MRSA, $I_h$ is the number of patients infected with HA-MRSA, $C_c$ is the number of patients colonized with CA-MRSA, $I_c$ is the number of patients infected with CA-MRSA, and we assume that patients who are placed in isolation stay in the removed state ($R$), so $R$ is the number of patients on additional precautions (i.e. isolation). The parameters $\beta_{CH}$, $\beta_{IH}$,  $\beta_{CC}$, $\beta_{IC}$ are the rates of new HA-MRSA colonization due to patients colonized with HA-MRSA, patients infected with HA-MRSA, patients colonized with CA-MRSA, and patients infected with CA-MRSA, correspondingly. Let $\sigma$ be the rate of new HA-MRSA colonization due to unobserved cases. Let $\alpha$ represent the rate of infection of patients colonized with HA-MRSA. We let $\rho_1$, $\rho_2$, $\rho_3$, and $\rho_4$ denote the inverse average time spent in hospitals before isolation for patients colonized with HA-MRSA, patients infected with HA-MRSA, patients colonized with CA-MRSA, and patients infected with CA-MRSA, respectively.}
  \label{fig:mrsamodel}
\end{figure}

\textbf{Model structure.} Figure \ref{fig:mrsamodel} describes a compartmental model representing the transmission dynamics of HA-MRSA and CA-MRSA in hospitals in Edmonton. The model comprises seven compartments: susceptible individuals ($S$), patients colonized with HA-MRSA ($C_h$), patients infected with HA-MRSA ($I_h$), patients colonized with CA-MRSA ($C_c$), patients infected with CA-MRSA ($I_c$), patients identified as having MRSA (including colonization and infection) who are removed from the MRSA transmission dynamics ($R$), such as those on additional precautions (i.e. isolation).

The arrows in the model (Figure \ref{fig:mrsamodel}) indicate the direction of transition of patients among these compartments. The number of susceptible individuals changes when new patients are admitted to or discharged from hospitals, and the number of patients newly colonized with HA-MRSA is influenced by the number of patients already colonized or infected with HA-MRSA ($C_h$, $I_h$) or CA-MRSA ($C_c$, $I_c$) \citep{cleveland}. Patients who have been colonized by either HA-MRSA ($C_h$) or CA-MRSA ($C_c$) may subsequently become infected, and these patients will be categorized as having infections with HA-MRSA ($I_h$) or CA-MRSA ($I_c$), respectively.  Patients colonized with HA-MRSA ($C_h$), patients infected with HA-MRSA ($I_h$), patients colonized with CA-MRSA ($C_c$) and patients infected with CA-MRSA ($I_c$) could be transferred to isolation rooms ($R$).

For susceptible individuals, the initial value ($S_0$) is 3048, which is the number of acute care beds in Edmonton in 2023 (internal communication). The initial values for patients colonized with HA-MRSA ($C_{h,0}$), patients infected with HA-MRSA ($I_{h,0}$), patients colonized with CA-MRSA ($C_{c,0}$), patients infected with CA-MRSA ($I_{c,0}$) are taken from the AHS IPC Edmonton data. We assume that there are no removed patients at the start of the study ($R_{0}=0$).

\textbf{Parameters.} The parameters correspond to the rates of dissemination among these compartments (Tables \ref{tab:beta} and \ref{tab:rho}). The transition rate of susceptible individuals ($S$) to patients colonized with HA-MRSA ($C_h$) is $\beta_{CH}C_h/N+\beta_{IH}I_h/N+\beta_{CC}C_c/N+\beta_{IC}I_c/N + \sigma$, where $\beta_{CH}$, $\beta_{IH}$, $\beta_{CC}$, $\beta_{IC}$ and $\sigma$ are the rates of colonization from patients colonized with HA-MRSA ($C_h$), patients infected with HA-MRSA ($I_h$), patients colonized with CA-MRSA ($C_c$), patients infected with CA-MRSA ($I_c$), and unobserved cases, respectively. The parameter $\alpha$ represents the rate of infection of patients colonized with HA-MRSA ($C_h$). Let $\rho_1$, $\rho_2$, $\rho_3$, and $\rho_4$ represent the average waiting time before isolation for patients with HA-MRSA colonization ($C_h$), patients with HA-MRSA infection ($I_h$), patients with CA-MRSA colonization ($C_c$), and patients with CA-MRSA infection ($I_c$), respectively. We assume here that patients who go into isolation enter the removed state ($R$) where they remain. The $R$ compartment essentially captures all individuals who were colonized or infected with MRSA in the study.

\begin{table}[htbp]
  \caption{Parameters and their priors in the MRSA transmission model.}
  \label{tab:beta}
  \centering
  \begin{tabular}{p{6.2cm} p{1.5cm} p{1.5cm} p{1.5cm} p{1.5cm}}    
    \hline \hline
    Parameter & Symbol  & Prior 1 & Prior 2 & Prior 3 \\
    \hline
    Transmission rate: the rate of new HA-MRSA colonization due to patients colonized with HA-MRSA. & $\beta_{CH}$ &  $\Gamma (1, 0.5)$ & $\Gamma (1, 1)$ & $\Gamma (1, 1.5)$\\ 
    \hline
    Transmission rate: the rate of new HA-MRSA colonization due to patients infected with HA-MRSA. & $\beta_{IH}$ & $\Gamma (1, 0.5)$ & $\Gamma (1, 1)$ & $\Gamma (1, 1.5)$ \\ 
    \hline
    Transmission rate: the rate of new HA-MRSA colonization due to patients colonized with CA-MRSA. & $\beta_{CC}$ & $\Gamma (1, 0.5)$ & $\Gamma (1, 1)$ & $\Gamma (1, 1.5)$ \\ 
    \hline
    Transmission rate: the rate of new HA-MRSA colonization due to patients infected with CA-MRSA. & $\beta_{IC}$ & $\Gamma (1, 0.5)$ & $\Gamma (1, 1)$ & $\Gamma (1, 1.5)$  \\
    \hline
    Import parameter: the rate of new HA-MRSA colonization due to unobserved cases. & $\sigma$ & $\Gamma (1, 0.5)$ & $\Gamma (1, 1)$ & $\Gamma (1, 1.5)$ \\ 
    \hline
    Infection rate: the rate of infection of HA-MRSA colonized patients. & $\alpha$ & $\Gamma (1, 0.5)$ & $\Gamma (1, 1)$ & $\Gamma (1, 1.5)$ \\ 
    \hline \hline
  \end{tabular}
\end{table}

\begin{table}[htbp]
  \caption{Parameters with values in the MRSA transmission model.}
  \label{tab:rho}
  \centering
  \begin{tabular}{p{6.2cm} p{1.5cm} p{1.5cm} p{3.5cm}}       
    \hline \hline
    Parameter & Symbol  & Value & Source \\
    \hline
    Inverse time to isolation: patients colonized with HA-MRSA stay in hospitals for an average time of $1/\rho_1$ before isolation. & $\rho_1$ & $1.3$ &\cite{loscanada} \\ 
    \hline
    Inverse time to isolation: patients infected with HA-MRSA stay in hospitals for an average time of $1/\rho_2$ before isolation. & $\rho_2$ & $1.3$ & \cite{loscanada} \\ 
    \hline
    Inverse time to isolation: patients colonized with CA-MRSA stay in hospitals for an average time of $1/\rho_3$ before isolation. & $\rho_3$ & $10$ & \cite{provsurv} \\ 
    \hline
    Inverse time to isolation: patients infected with CA-MRSA stay in hospitals for an average time of $1/\rho_4$ before isolation. & $\rho_4$ & $10$ & \cite{provsurv} \\
    \hline \hline
  \end{tabular}
\end{table}

\textbf{Transmission dynamics.} Let $t\in \{1, \dots, \tau\}$ denote the observed time during the research period, and $ \tau=61$ because there are $61$ months between September 2018 and September 2023. At the time $t$, $S_t$, $C_{h,t}$, $I_{h,t}$, $C_{c,t}$, $I_{c,t}$ and $R_t$ represent the number of individuals susceptible, colonized with HA-MRSA, infected with HA-MRSA, colonized with CA-MRSA, infected with CA-MRSA, and removed, respectively. The total number of participants in the study population at time $t$ includes all compartments, which is $N_t=S_{t}+C_{h,t}+I_{h,t}+C_{c,t}+I_{c,t}+R_{t}$. Let $C^*_{h,t}$, $I^*_{h,t}$, $C^*_{c,t}$, $I^*_{c,t}$ represent the number of people newly colonized with HA-MRSA, infected with HA-MRSA, colonized with CA-MRSA and infected with CA-MRSA at time $t$, respectively. Let $R^*_{1,t}$, $R^*_{2,t}$, $R^*_{3,t}$ and $R^*_{4,t}$ represent the number of persons colonized with HA-MRSA, infected with HA-MRSA, colonized with CA-MRSA and infected with CA-MRSA who are placed on isolation in hospitals. 

At time $t+1$, patients admitted to and discharged from hospitals at time $t$ ($\text{Admission}_t$ and $\text{Discharge}_t$) could increase or decrease the number of susceptible individuals $S_{t+1}$, as well as the total number of individuals in the model $N_{t+1}$. The difference equations describing the flow through the compartments are:
\begin{align*} 
  &S_{t+1}=S_t+\text{Admission}_t-\text{Discharge}_t-C^*_{h,t}, \\   
  &C_{h,t+1}=C_{h,t}+C^*_{h,t}-I^*_{h,t}-R^*_{1,t} ,\\
  &I_{h,t+1}=I_{h,t}+I^*_{h,t}-R^*_{2,t}, \\
  &C_{c,t+1}=C_{c,t}+C^*_{c,t}-R^*_{3,t}, \\
  &I_{c,t+1}=I_{c,t}+I^*_{c,t}-R^*_{4,t}, \\
  &R_{t+1}=R_t+R^*_{1,t}+R^*_{2,t}+R^*_{3,t}+R^*_{4,t}.
\end{align*}

Let $\pi^{(SC_h)}_t$ denote the probability that a patient colonized or infected with MRSA transmits the bacterium to a susceptible individual in hospitals, $\pi^{(C_hI_h)}$ represent the probability of a patient colonized with HA-MRSA becoming infected, $\pi^{(C_hR)}$, $\pi^{(I_hR)}$, $\pi^{(C_cR)}$ and $\pi^{(I_cR)}$ are the transition probabilities for isolating patients colonized with HA-MRSA, infected with HA-MRSA, colonized with CA-MRSA, and infected with CA-MRSA, respectively. The probabilities $\pi^{(SC_h)}_t$, $\pi^{(C_hI_h)}$,  $\pi^{(C_hR)}$, $\pi^{(I_hR)}$, $\pi^{(C_cR)}$ and $\pi^{(I_cR)}$ are specified as follows:
\begin{align*}
  &\pi^{(SC_h)}_t = 1-\text{exp}\bigg[-\bigg(\frac{\beta_{CH}C_{h,t}}{N_t}+\frac{\beta_{IH}I_{h,t}}{N_t}+\frac{\beta_{CC}C_{c,t}}{N_t}+\frac{\beta_{IC}I_{c,t}}{N_t}+\sigma \bigg) \bigg],\\
  &\pi^{(C_hI_h)} = 1-\text{exp}(-\alpha),\\
  &\pi^{(C_hR)} = 1-\text{exp}(-\rho_1),\\
  &\pi^{(I_hR)} = 1-\text{exp}(-\rho_2),\\
  &\pi^{(C_cR)} = 1-\text{exp}(-\rho_3),\\
  &\pi^{(I_cR)} = 1-\text{exp}(-\rho_4),
\end{align*}
where the definitions of parameters $\beta_{CH}$, $\beta_{IH}$, $\beta_{CC}$, $\beta_{IC}$, $\sigma$, $\alpha$, $\rho_1$, $\rho_2$, $\rho_3$, and $\rho_4$ are described in Tables \ref{tab:beta} and \ref{tab:rho}. In order to ascertain how quickly various groups of MRSA patients are transferred to isolated rooms, we choose $\rho_1=\rho_2=1.3$ because we assume a three-week timeframe for patients admitted to hospitals, colonized or infected with HA-MRSA, and then isolated after being diagnosed with MRSA \citep{loscanada}. In the meantime, patients colonized or infected with CA-MRSA are often shifted to isolated wards on the third day of admission, hence $\rho_3$ and $\rho_4$ are set to $10$  \citep{provsurv}. The transitions between compartments are assumed to follow binomial distributions: 
\begin{align*}
  &C^*_{h,t} \sim \text{Bin} \big(S_t, \pi^{(SC_h)}_t \big), \\
  &I^*_{h,t} \sim \text{Bin} \big(C_{h,t}, \pi^{(C_hI_h)} \big), \\
  &R^*_{1,t} \sim \text{Bin} \big(C_{h,t}, \pi^{(C_hR)} \big), \\
  &R^*_{2,t} \sim \text{Bin} \big(I_{h,t}, \pi^{(I_hR)} \big), \\
  &R^*_{3,t} \sim \text{Bin} \big(C_{c,t}, \pi^{(C_cR)} \big), \\
  &R^*_{4,t} \sim \text{Bin} \big(I_{c,t}, \pi^{(I_cR)} \big). 
\end{align*} 

At the time $t$, patients newly colonized with HA-MRSA $C^*_{h,t}$ are the number of susceptible people colonized with HA-MRSA with probability $\pi^{(SC_h)}_t$, patients newly infected with HA-MRSA $I^*_{h,t}$ are the number of patients colonized with HA-MRSA who develop infections at the rate $\pi^{(C_hI_h)}$. Patients colonized with HA-MRSA, infected with HA-MRSA, colonized with CA-MRSA, and infected with CA-MRSA are transferred to isolated wards with transition probabilities $\pi^{(C_hR)}$, $\pi^{(I_hR)}$, $\pi^{(C_cR)}$, and $\pi^{(I_cR)}$, respectively. Therefore, the log-likelihood function of the chain binomial MRSA transmission model is
\begin{align*}
  \ell({D} | {\Theta}, {\Omega}) =& \sum_{t=1}^ \tau \bigg [ \text{log} \binom {S_t}{C^*_{h,t}} + C^*_{h,t} \text{log}\pi^{(SC_h)}_t + \big(S_t-C^*_{h,t}\big) \text{log} \big(1-\pi^{(SC_h)}_t \big) 
\\ & +  \text{log} \binom {C_{h,t}}{I^*_{h,t}} + I^*_{h,t}\text{log}\pi^{(C_hI_h)} + \big(C_{h,t}-I^*_{h,t}\big)  \text{log} \big(1-\pi^{(C_hI_h)} \big)
\\ & +  \text{log} \binom {C_{h,t}}{R^*_{1,t}} + R^*_{1,t}\text{log}\pi^{(C_hR)} + \big(C_{h,t}-R^*_{1,t}\big)  \text{log} \big(1-\pi^{(C_hR)}\big)
\\ & +  \text{log} \binom {I_{h,t}}{R^*_{2,t}} + R^*_{2,t}\text{log}\pi^{(I_hR)} + \big(I_{h,t}-R^*_{2,t}\big)  \text{log} \big(1-\pi^{(I_hR)} \big)
\\ & +  \text{log} \binom {C_{c,t}}{R^*_{3,t}} + R^*_{3,t}\text{log}\pi^{(C_cR)} + \big(C_{c,t}-R^*_{3,t}\big)  \text{log} \big(1-\pi^{(C_cR)}\big)
\\ & +  \text{log} \binom {I_{c,t}}{R^*_{4,t}} + R^*_{4,t}\text{log}\pi^{(I_cR)} + \big(I_{c,t}-R^*_{4,t}\big)  \text{log} \big(1-\pi^{(I_cR)}\big)
\bigg ],
\end{align*}
where ${\Theta} = (\beta_{CH}, \beta_{IH}, \beta_{CC}, \beta_{IC}, \sigma,  \alpha)^{\intercal}$ denote the vector of parameters to be estimated in the compartmental model (Table \ref{tab:beta}), ${\Omega} = (\rho_1, \rho_2, \rho_3, \rho_4)^{\intercal}$ denote the vector of parameters in the model with a specific value (Table \ref{tab:rho}). Let ${D}=(t_1,\dots,t_{\tau}, S_0, C_{h,0}, I_{h,0}, C_{c,0}, I_{c,0}, R_0, C^*_{h,t}, I^*_{h,t}, C^*_{c,t}, I^*_{c,t})$ denote the observed data. The initial values of the compartments in the model are $S_0$, $C_{h,0}$, $I_{h,0}$, $C_{c,0}$, $I_{c,0}$, and $R_0$. The number of susceptible individuals at the beginning of the study ($S_0$) is determined by the number of acute care beds in Edmonton. The start value of patients colonized with HA-MRSA ($C_{h,0}$), patients infected with HA-MRSA ($I_{h,0}$), patients colonized with CA-MRSA ($C_{c,0}$) and patients infected CA-MRSA ($I_{c,0}$) is the number of patients newly colonized with HA-MRSA, patients newly infected with HA-MRSA, patients newly colonized with CA-MRSA, and patients newly infected with CA-MRSA during the first month of the study period. $R_0$ is zero because it is presumed that there were no removed patients at the beginning.

\section{Bayesian inference and Markov chain Monte Carlo (MCMC) methods \label{sec:method}}
After developing a model for MRSA transmission in hospitals, we need to consider model fitting to data and parameter estimation. As the number of compartments in the model grows, so does the dimension of the parameter vector, making parameter estimation more difficult. To address this issue, we estimate parameters in our compartmental model using the Markov chain Monte Carlo (MCMC) method in conjunction with a Bayesian framework. 

In Bayesian inference, all parameters in the model are assumed to be random variables with specific distributions. The prior distribution $P({\Theta})$ reflects our initial beliefs and uncertainty about the parameters before observing the data ${D}$. In this study, we use weakly informative gamma priors to incorporate our belief about the parameter of interest into compartmental modelling (Table \ref{tab:beta}). To make the posterior more dependent on the likelihood function, we select weakly informative priors, whose means are greater than the estimated parameters. These prior settings are considered to test the sensitivity of results to prior assumptions. The posterior distribution $P({\Theta}|{D})$ represents our updated belief about the parameters after considering the observed data, which is proportional to the product of the prior distribution $P({\Theta})$ and the likelihood function of the model $\mathcal{L}({D}|{\Theta})$ \citep[Chapter 1]{bayesdata}. Symbolically, we can express this as: $P({\Theta}|{D}) \propto \mathcal{L}({D}|{\Theta})P({\Theta})$, where $\ell({D}|{\Theta}) = \text{log}[\mathcal{L}({D}|{\Theta})]$.

MCMC is a computational technique that draws samples from the posterior distribution \citep{introMCMC}. This efficient method allows us to obtain estimated parameter values by calculating the sample mean, which has been widely used in compartmental models \citep{stocMCMC, bayeshainfection}. To implement MCMC, acceptance criteria are established to determine whether each draw should be accepted or rejected based on the posterior distribution. As the algorithm iterates, a chain of accepted draws is produced, which eventually converges to the posterior distribution. We discard the initial burn-in of 10,000 iterations (from 60,000) to ensure the chain reaches the stationary distribution. We employ a random walk Metropolis-Hastings (RWMH) algorithm \citep[Chapter 6]{MCMCR} as it allows for step-by-step gathering of information about the posterior distribution. We use the {R} \citep[v4.4.1]{citer} package \texttt{nimble} \citep{nimble, nimbleurl, nimblemanual} to implement the MCMC and model. The trace plots produced by the MCMC algorithm are used to assess the mixing of a chain. 

We first fit the full model to the data and find posterior estimates of the parameters using the MCMC algorithm. However, we also want to know which patient groups have a greater impact on MRSA transmission. Thus, we also fit a series of sub-models that make various assumptions regarding the sources of new MRSA colonization. Specifically, we consider models in which transmission risk depends on those who have been colonized with HA-MRSA, infected with HA-MRSA, colonized with CA-MRSA, or infected with CA-MRSA, and various combinations of these. 

To do this we modify the probability of a susceptible individual becoming colonized with MRSA, $\pi^{(SC_h)}_t$. The various sub-models considered are shown in Table \ref{tab:assumption}. For example, in Model 1, we assume that patients colonized with HA-MRSA ($C_h$) are the primary reason for new patients colonized with HA-MRSA in hospitals, and that there is no relationship between the other three patient groups ($I_h$, $C_c$, and $I_c$) and MRSA transmission. Hence, we only set $\beta_{IH}=\beta_{CC}=\beta_{IC}=0$, leading to $\pi^{(SC_h)}_t = 1-\text{exp}[-(\beta_{CH}C_{h,t}/N_t+\sigma)]$. Similarly, Models 2-4 take into account MRSA transmission from patients infected with HA-MRSA, patients colonized with CA-MRSA, and patients infected with CA-MRSA, respectively. After investigating the impact of a single group of patients on MRSA transmission in hospitals, the model could be expanded to include two groups of patients at the same time (Models 5-10), followed by three groups of patients (Models 11-14). Model 15 is the full model (Table \ref{fig:mrsamodel}), and it takes into account transmission for all four types of patients. 

We compare the models using the Widely Applicable Information Criteria (WAIC) \citep{waic2010}. WAIC is a useful method for assessing the predictive accuracy of a Bayesian model, and a lower WAIC usually indicates better model performance \citep{waic2017}. 

\begin{table}[htbp]
  \caption{Different assumptions for the MRSA transmission model.}
  \label{tab:assumption}
  \centering
  \begin{tabular}{p{1cm} p{8cm} p{3.5cm}}   
    \hline \hline  
    Model & Assumption on the source of new HA-MRSA colonization & Expression in the model\\
    \hline
    1 & Patients colonized with HA-MRSA only&  $\beta_{IH} = \beta_{CC} = \beta_{IC}= 0$\\ 
    \hline
    2 & Patients infected with HA-MRSA only&  $\beta_{CH} = \beta_{CC} = \beta_{IC}= 0$\\
    \hline
    3 & Patients colonized with CA-MRSA only&  $\beta_{CH} = \beta_{IH} = \beta_{IC}= 0$\\ 
    \hline
    4 & Patients infected with CA-MRSA only&  $\beta_{CH} = \beta_{IH} = \beta_{CC}= 0$\\ 
    \hline 
    5 & Patients colonized with HA-MRSA; Patients infected with HA-MRSA &  $\beta_{CC} = \beta_{IC}= 0$\\ 
    \hline     
    6 & Patients colonized with CA-MRSA; Patients infected with CA-MRSA &  $\beta_{CH} = \beta_{IH}= 0$\\ 
    \hline     
    7 & Patients colonized with HA-MRSA; Patients colonized with CA-MRSA &  $\beta_{IH} = \beta_{IC}= 0$\\ 
    \hline 
    8 & Patients infected with HA-MRSA; Patients infected with CA-MRSA &  $\beta_{CH} = \beta_{CC}= 0$\\ 
    \hline 
    9 & Patients colonized with HA-MRSA; Patients infected with CA-MRSA &  $\beta_{IH} = \beta_{CC}= 0$\\ 
    \hline 
    10 & Patients infected with HA-MRSA; Patients colonized with CA-MRSA &  $\beta_{CH} = \beta_{IC}= 0$\\ 
    \hline 
    11 & Patients colonized with HA-MRSA; Patients infected with HA-MRSA; Patients colonized with CA-MRSA &  $\beta_{IC}= 0$\\ 
    \hline 
    12 & Patients colonized with HA-MRSA; Patients colonized with CA-MRSA; Patients infected with CA-MRSA  &  $\beta_{IH}= 0$\\ 
    \hline
    13 & Patients colonized with HA-MRSA; Patients infected with HA-MRSA; Patients infected with CA-MRSA  &  $\beta_{CC}= 0$\\ 
    \hline 
    14 & Patients infected with HA-MRSA; Patients colonized with CA-MRSA; Patients infected with CA-MRSA  &  $\beta_{CH}= 0$\\ 
    \hline 
    15 & Patients colonized with HA-MRSA; Patients infected with HA-MRSA; Patients colonized with CA-MRSA; Patients infected with CA-MRSA  &  /\\ 
    \hline \hline
  \end{tabular}
\end{table}

\section{Results \label{sec:result}}
\begin{sidewaystable}[p]
  \caption{MRSA transmission model comparison based on various assumptions about the source of new MRSA colonization in hospitals of Edmonton.} 
  \label{tab:comparison}
  \centering
  \begin{threeparttable}
  \begin{tabular}{ccccccccccc}
    \hline \hline
    \multicolumn{3}{c}{Prior 1: $\Gamma (1, 0.5)$} & \multicolumn{3}{c}{Prior 2: $\Gamma (1, 1)$} & \multicolumn{3}{c}{Prior 3: $\Gamma (1, 1.5)$} \\   
    \cline{1-3} \cline{4-6} \cline{7-9}
    Model & Parameter assumption \tnote{a} & WAIC \tnote{b} & Model & Parameter assumption \tnote{a} & WAIC \tnote{b}  & Model & Parameter assumption \tnote{a} & WAIC \tnote{b} \\
    \hline
    4 & $\beta_{IC}$ & 888.10 & 2 & $\beta_{IH}$ & 888.03 & 1 & $\beta_{CH}$ & 888.16 \\ 
    1 & $\beta_{CH}$ & 888.12 & 4 & $\beta_{IC}$ & 888.10 & 2 & $\beta_{IH}$ & 888.19 \\ 
    3 & $\beta_{CC}$ & 888.15 & 3 & $\beta_{CC}$ & 888.14 & 4 & $\beta_{IC}$ & 888.22 \\ 
    2 & $\beta_{IH}$ & 888.18 & 1 & $\beta_{CH}$ & 888.22 & 3 & $\beta_{CC}$ & 888.32 \\ 
    10 & $\beta_{IH}$, $\beta_{CC}$ & 889.98 & 5 &  $\beta_{CH}$, $\beta_{IH}$ & 890.13 & 8 & $\beta_{IH}$, $\beta_{IC}$ & 890.16 \\ 
    8 & $\beta_{IH}$, $\beta_{IC}$ & 889.99 & 10 & $\beta_{IH}$, $\beta_{CC}$ & 890.13 & 6 & $\beta_{CC}$, $\beta_{IC}$  & 890.18 \\ 
    9 & $\beta_{CH}$, $\beta_{IC}$ & 890.01 & 9 & $\beta_{CH}$, $\beta_{IC}$ & 890.23 & 10 & $\beta_{IH}$, $\beta_{CC}$ & 890.23 \\ 
    5 & $\beta_{CH}$, $\beta_{IH}$ & 890.13 & 6 & $\beta_{CC}$, $\beta_{IC}$ & 890.24 & 5 &  $\beta_{CH}$, $\beta_{IH}$ & 890.30 \\ 
    6 &  $\beta_{CC}$, $\beta_{IC}$ & 890.19 & 7 & $\beta_{CH}$, $\beta_{CC}$ & 890.28 & 7 & $\beta_{CH}$, $\beta_{CC}$ & 890.34 \\ 
    7 & $\beta_{CH}$, $\beta_{CC}$  & 890.25 & 8 & $\beta_{IH}$, $\beta_{IC}$ & 890.35 & 9 & $\beta_{CH}$, $\beta_{IC}$ & 890.45 \\ 
    13 & $\beta_{CH}$, $\beta_{IH}$, $\beta_{IC}$ & 891.81 & 12 & $\beta_{CH}$, $\beta_{CC}$, $\beta_{IC}$ & 892.06 & 14 & $\beta_{IH}$, $\beta_{CC}$, $\beta_{IC}$ & 892.19 \\ 
    12 & $\beta_{CH}$, $\beta_{CC}$, $\beta_{IC}$ & 891.87 & 14 & $\beta_{IH}$, $\beta_{CC}$, $\beta_{IC}$ & 892.08 & 12 & $\beta_{CH}$, $\beta_{CC}$, $\beta_{IC}$ & 892.26 \\ 
    14 & $\beta_{IH}$, $\beta_{CC}$, $\beta_{IC}$ & 891.95 & 11 & $\beta_{CH}$, $\beta_{IH}$, $\beta_{CC}$ & 892.08 & 11 & $\beta_{CH}$, $\beta_{IH}$, $\beta_{CC}$ & 892.31 \\ 
    11 & $\beta_{CH}$, $\beta_{IH}$, $\beta_{CC}$ & 891.98 & 13 & $\beta_{CH}$, $\beta_{IH}$, $\beta_{IC}$ & 892.22 & 13 & $\beta_{CH}$, $\beta_{IH}$, $\beta_{IC}$ & 892.46 \\ 
   15 & $\beta_{CH}$, $\beta_{IH}$, $\beta_{CC}$, $\beta_{IC}$ & 893.95 & 15 & $\beta_{CH}$, $\beta_{IH}$, $\beta_{CC}$, $\beta_{IC}$ & 894.11 & 15 & $\beta_{CH}$, $\beta_{IH}$, $\beta_{CC}$, $\beta_{IC}$ & 894.07 \\ 
    \hline
    S.d.\tnote{c}   &  & 1.75 &  &  & 1.82 &  &  & 1.83 \\    
    \hline \hline
  \end{tabular}
  \begin{tablenotes}
    \item[a] This indicates that these parameters are included in the models.
    \item[b] Widely Applicable Information Criteria.
    \item[c] Standard deviation.
  \end{tablenotes}
  \end{threeparttable}
\end{sidewaystable}

In Table \ref{tab:comparison}, we compare the WAIC results for the fifteen models fitted to the data under these prior assumptions. These models are presented in ascending order of WAIC value, from lowest to highest. We found that when these models were fitted to the data with prior 1 ($\Gamma (1, 0.5)$), they could be divided into four groups based on their WAIC. The first four models (Models 4, 1, 3 and 2) in Group A have WAIC values ranging from 888.10 to 888.18. Because the difference between the lowest WAIC and the fourth lowest WAIC is less than one, we can conclude that these models perform approximately equally well. This observation suggests that all patient groups, including those colonized with HA-MRSA ($C_h$), infected with HA-MRSA ($I_h$), colonized with CA-MRSA ($C_c$), and infected with CA-MRSA ($I_c$), have an impact on MRSA transmission in hospitals. 

Group B includes the following six model assumptions (Models 10, 8, 9, 5, 6 and 7), with WAIC values ranging from 889.98 to 890.25, indicating that these six models perform similarly due to minor differences in WAIC. The impact of MRSA transmission from two patient groups is examined by Group B in comparison to the models in Group A. We find that the WAIC does increase as more parameters are entered into the transmission risk function, suggesting penalization for complexity. Another example of this phenomenon is seen in Group C's WAIC value (Models 13, 12, 14 and 11), which ranges from 891.81 to 891.98, where adding a parameter increases the WAIC value by about one, and by two in Model 15 (Group D), which is the full model of MRSA transmission. Since the WAIC value increased by 5 points for Model 15 compared to the models in Group A, we can still choose any of these models to fit the data, even though the WAIC value increases slightly as the number of parameters in the models increases. With a maximum difference of 5-6 in the WAICs, there is no evidence that any of the models fit better than any others. Thus, it is difficult to identify which patient group has a higher impact of MRSA transmission in Edmonton's hospitals.

We also assume the other two weakly informative priors, prior 2 ($\Gamma (1, 1)$) and prior 3 ($\Gamma (1, 1.5)$), shown in the second column and the third column in Table \ref{tab:comparison}. We find that priors have negligible effects on WAICs, so the conclusion according to WAICs stays broadly unchanged, demonstrating that our findings are robust to the choice of priors.

\begin{sidewaystable}[p]
  \caption{The estimated results for parameters in the full MRSA transmission model (Model 15). Model 15 assumes that patients colonized with HA-MRSA ($C_h$), infected with HA-MRSA ($I_h$), colonized with CA-MRSA ($C_c$), and infected with CA-MRSA ($I_c$) all influence MRSA transmission in hospitals.} 
  \label{tab:fullmodel}
  \centering
  \begin{threeparttable}
  \begin{tabular}{cccccccccccc}
    \hline \hline
    \multicolumn{3}{c}{Prior 1: $\Gamma (1, 0.5)$} & \multicolumn{3}{c}{Prior 2: $\Gamma (1, 1)$} & \multicolumn{3}{c}{Prior 3: $\Gamma (1, 1.5)$} \\
    \cline{1-3} \cline{4-6}\cline{7-9}
    Parameter & Posterior mean & 95\% C.I.\tnote{a} & Parameter & Posterior mean & 95\% C.I.\tnote{a} & Parameter & Posterior mean & 95\% C.I.\tnote{a} \\
    \hline
    $\beta_{CH}$ & 0.0421 & (0.0011, 0.1494) & $\beta_{CH}$ & 0.0439 & (0.0010, 0.1614) & $\beta_{CH}$ & 0.0436 & (0.0011, 0.1558) \\  
    $\beta_{IH}$ & 0.0567 & (0.0015, 0.2036) & $\beta_{IH}$ & 0.0612 & (0.0017, 0.2190) & $\beta_{IH}$ & 0.0622 & (0.0017, 0.2213)  \\   
    $\beta_{CC}$ & 0.0095 & (0.0002, 0.0350) & $\beta_{CC}$ & 0.0094 & (0.0002, 0.0341) & $\beta_{CC}$ & 0.0095 & (0.0003, 0.0342) \\   
    $\beta_{IC}$ & 0.0407 & (0.0011, 0.1500) & $\beta_{IC}$ & 0.0399 & (0.0011, 0.1400) & $\beta_{IC}$ & 0.0417 & (0.0010, 0.1500) \\     
    $\sigma$ & 0.0100 & (0.0091, 0.0107) &$\sigma$ & 0.0100 & (0.0091, 0.0107) & $\sigma$ & 0.0100 & (0.0091, 0.0107) \\ 
    $\alpha$ & 0.2628 & (0.2358, 0.2910) & $\alpha$ & 0.2630 & (0.2361, 0.2910) & $\alpha$ & 0.2631 & (0.2369, 0.2911) \\ 
    \hline \hline
  \end{tabular}
  \begin{tablenotes}
    \item[a] Credible Interval.
  \end{tablenotes}
  \end{threeparttable}
\vspace{0.5in}
  \caption{The estimated results for parameters in the subset MRSA transmission models (Models 4, 2 and 1). Model 4  assumes that only patients infected with CA-MRSA ($I_c$) influence MRSA transmission in hospitals. Model 2  assumes that only patients infected with HA-MRSA ($I_h$) influence MRSA transmission in hospitals. Model 1  assumes that only patients colonized with HA-MRSA ($C_h$) influence MRSA transmission in hospitals.}
  \label{tab:submodel}
  \centering
  \begin{threeparttable}
  \begin{tabular}{cccccccccccc}
    \hline \hline
    \multicolumn{3}{c}{Model 4 with Prior 1: $\Gamma (1, 0.5)$} & \multicolumn{3}{c}{Model 2 with Prior 2: $\Gamma (1, 1)$} & \multicolumn{3}{c}{Model 1 with Prior 3: $\Gamma (1, 1.5)$} \\
    \cline{1-3} \cline{4-6} \cline{7-9}
    Parameter & Posterior mean & 95\% C.I.\tnote{a} & Parameter & Posterior mean & 95\% C.I.\tnote{a} & Parameter & Posterior mean & 95\% C.I.\tnote{a} \\
    \hline  
    $\beta_{IC}$ & 0.0420 & (0.0009, 0.1538) & $\beta_{IH}$ & 0.0621 & (0.0017, 0.2204) & $\beta_{CH}$ & 0.0441 & (0.0012, 0.1567) \\      
    $\sigma$ & 0.0104 & (0.0097, 0.0110) &$\sigma$ & 0.0105 & (0.0099, 0.0110) & $\sigma$ & 0.0103 & (0.0096, 0.0110) \\ 
    $\alpha$ & 0.2633 & (0.2372, 0.2906) & $\alpha$ & 0.2633 & (0.2366, 0.2910) & $\alpha$ & 0.2633 & (0.2372, 0.2909) \\  
    \hline \hline
  \end{tabular}
  \begin{tablenotes}
    \item[a] Credible Interval.
  \end{tablenotes}
  \end{threeparttable}
\end{sidewaystable}

Since all fifteen models fit the data fairly evenly based on their similar WAIC values, we further explored the full model (Model 15) and the simple models that produced the lowest WAIC under each prior (Models 1, 2 and 4). Model 15 assumes that patients colonized or infected with HA-MRSA or CA-MRSA affect the spread of MRSA in hospitals, and contains all parameters of ${\Theta}$, where ${\Theta} = (\beta_{CH}, \beta_{IH}, \beta_{CC}, \beta_{IC}, \sigma, \alpha)^{\intercal}$ in the analysis. 

The rate of transmission due to patients colonized with HA-MRSA ($\beta_{CH}$) has a posterior mean of 0.04, meaning 0.04 susceptible individuals are colonized with HA-MRSA by each patient colonized with HA-MRSA per month (Table \ref{tab:fullmodel}). Multiplying $\beta_{CH}$ by the mean of the total number of patients colonized with HA-MRSA in hospitals over the study period, which is 704, we can estimate that approximately 28 people with a 95\% C.I. (0.7, 105) are colonized with HA-MRSA every month by patients already colonized with HA-MRSA in hospitals. 

Similarly, $\beta_{IH}$ is the rate of new HA-MRSA colonization due to patients infected with HA-MRSA, with approximately 0.06 susceptible individuals colonized with HA-MRSA per contact per month. Based on the average number of patients infected with HA-MRSA in hospitals (177), we can estimate that patients infected with HA-MRSA in hospitals colonize around 11 individuals per month with a 95\% C.I. (0.3, 36).

Recall that $\beta_{CC}$ represents the rate of new HA-MRSA colonization in hospitals due to contact between susceptible individuals and patients colonized with CA-MRSA, and we calculate that approximately 13 susceptible individuals with a 95\% C.I. (0.3, 51) are colonized with HA-MRSA each month by patients colonized with CA-MRSA, since the mean of total patients colonized with CA-MRSA in hospitals during the study period is 1471. This means that one patient colonized with CA-MRSA colonizes an average of around 0.009 susceptible individuals each month under the posterior mean. 

The posterior mean estimate of $\beta_{IC}$ suggests that one patient infected with CA-MRSA colonizes around 0.04 susceptible people per contact per month. Because the average number of individuals infected with CA-MRSA is 669, approximately 27 susceptible people with a 95\% C.I. (0.7, 100) are colonized with HA-MRSA each month in hospitals due to patients infected with CA-MRSA.

The transmission rate from the susceptible group to the HA-MRSA colonization group due to unobserved patients is denoted by $\sigma$, and one latent MRSA case per contact per month colonizes an average of about 0.01 susceptible persons under the posterior mean (95 \% C.I.: (0.009, 0.011)). In addition, HA-MRSA colonized patients have an infection rate ($\alpha$) of around 0.26 with a 95\% C.I. (0.24, 0.29).

The transmission rate from the susceptible group to the HA-MRSA colonization group due to unobserved patients is denoted by $\sigma$, and one latent MRSA case per contact per month colonizes an average of about 0.01 susceptible persons under the posterior mean. In addition, HA-MRSA colonized patients have an infection rate ($\alpha$) of around 0.26.

We depict the posterior estimation of parameters for subset models which has the lowest WAIC value (Models 4, 2 and 1) using different priors in Table \ref{tab:submodel}. Models 1, 2 and 4 focus primarily on the impact of patients colonized with HA-MRSA, infected with HA-MRSA, and infected with CA-MRSA on MRSA spread, using the transmission rates $\beta_{CH}$, $\beta_{IH}$, and $\beta_{IC}$, repectively. When comparing the posterior mean of these parameters with these under the full model, their estimated values are similar (Table \ref{tab:fullmodel}). 

\begin{figure}[htbp]
  \centering
  \includegraphics[width=16cm]{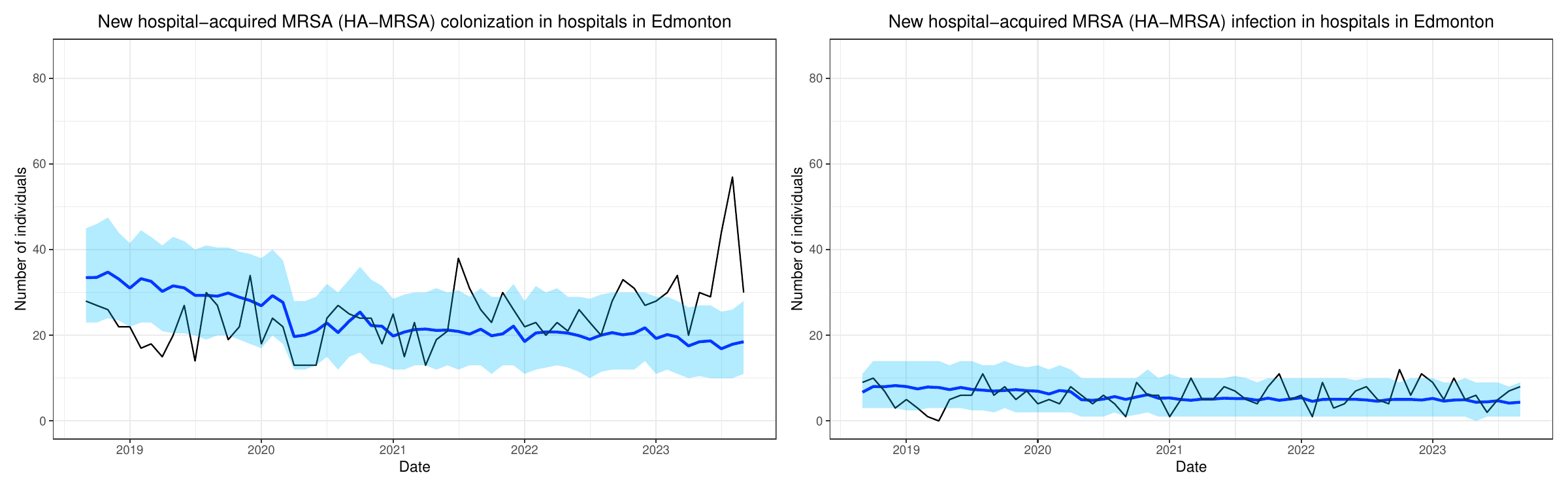}
  \caption{Simulated data from the full MRSA transmission model (Model 15) using prior 1 ($\Gamma (1, 0.5)$) and real data from hospitals in Edmonton for new HA-MRSA colonization (left) (MAE: 7.63) and infection (right) (MAE: 2.36). The black line represents real data, the deep blue line represents the mean of simulated data for each month, and the shades of sky blue in the upper and lower bounds represent 95\% simulated data.}
  \label{fig:edmonton15}
\end{figure}

\begin{figure}[htbp]
  \centering
  \includegraphics[width=16cm]{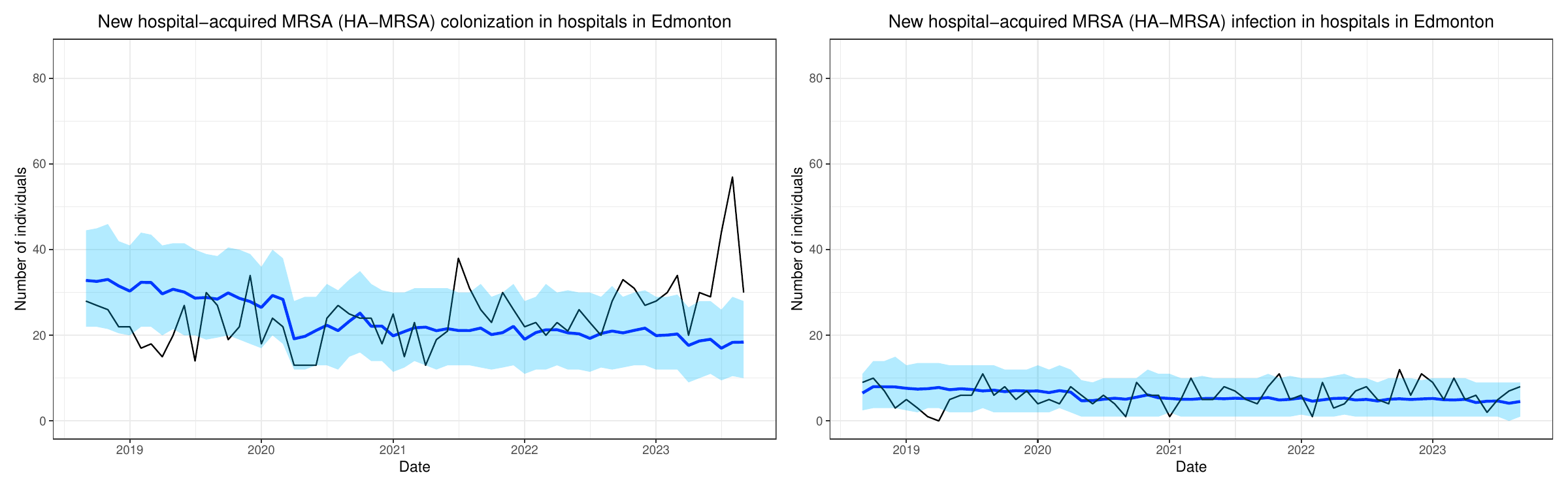}
  \caption{Simulated data from the subset MRSA transmission model (Model 4) using prior 1 ($\Gamma (1, 0.5)$) and real data from hospitals in Edmonton for new HA-MRSA colonization (left) (MAE: 7.36) and infection (right) (MAE: 2.35). The black line represents real data, the deep blue line represents the mean of simulated data for each month, and the shades of sky blue in the upper and lower bounds represent 95\% simulated data.}
  \label{fig:edmonton4}
\end{figure}

We compare simulated data from the MRSA transmission model under the posterior distribution (Model 15 and Model 4) using prior 1 ($\Gamma (1, 0.5)$) with actual data from hospitals in Edmonton in Figures \ref{fig:edmonton15} to \ref{fig:edmonton4}. Because the plots using prior 2 ($\Gamma (1, 1)$) and prior 3 ($\Gamma (1, 1.5)$) are very similar to prior 1, we only provide the simulated results with prior 1. Both models show 95\% overlap with real data on most months, indicating a good fit between September 2018 and June 2023. However, Model 15 and Model 4 failed to capture the large spike from July 2023 to September 2023 (near the end of the study), indicating that these models may be unable to capture certain unexpected increases in cases, or potential outbreaks that occurred in real-world scenarios. The simulated data of new patients infected with HA-MRSA (on the right side in Figures \ref{fig:edmonton15} and \ref{fig:edmonton4}) match the real data for almost the entire study period, and their 95\% interval overlaps with the actual data. The mean absolute error (MAE) computes the average absolute difference between the actual and simulated data. New patients infected with HA-MRSA have a lower MAE compared to new patients colonized with HA-MRSA, indicating that these models have higher predictive accuracy on the number of patients newly infected with HA-MRSA in hospitals in Edmonton, compared to the number of patients newly colonized with HA-MRSA.

\section{Discussion \label{sec:discussion}}
MRSA is prevalent in hospitals, can easily spread, can lead to infections, and can result in high morbidity and mortality among patients hospitalized in acute care facilities in Canada \citep{sysreview}. To investigate MRSA dissemination in hospitals, we developed a stochastic compartmental model with seven compartments, which are susceptible individuals, patients colonized with HA-MRSA, patients infected with HA-MRSA, patients colonized with CA-MRSA, patients infected with CA-MRSA, and patients identified with MRSA who are isolated. 

Then, we investigated how patients migrate between these compartments by estimating transmission rates, exhibiting MRSA spread in hospitals in Edmonton. In the model, we assumed that the number of susceptible individuals fluctuates as new patients are admitted to or discharged from hospitals. The number of patients who have already been colonized or infected with HA-MRSA or CA-MRSA in hospitals influences the number of patients who are newly colonized with HA-MRSA because we assumed MRSA spreads through direct contact between susceptible people in hospitals and these patient populations. Meanwhile, patients may later contract an infection from either HA-MRSA or CA-MRSA colonization, and then these patients are classified as having HA-MRSA or CA-MRSA infections, respectively. Furthermore, patients who have been colonized with HA-MRSA, infected with HA-MRSA, colonized with CA-MRSA, or infected with CA-MRSA may be isolated.

We first fit the full model to the patient data from hospitals in Edmonton using a Bayesian framework, and then we estimated the posterior mean using MCMC techniques. Next, we changed the model's assumptions to assume that the new MRSA colonization in the hospital is the result of direct effective contact between susceptible individuals and one or more of the patient groups that were previously colonized with HA-MRSA, infected with HA-MRSA, colonized with CA-MRSA, or infected with CA-MRSA. This allows us to investigate which patient categories have the most impact on MRSA transmission in hospitals. 

Under the full model, we observed that approximately 28 (95\% C.I.: (0.7, 105)) susceptible individuals are colonized with HA-MRSA by patients colonized with HA-MRSA each month, and approximately 27  (95\% C.I.: (0.7, 100)) people are colonized with HA-MRSA by patients infected with CA-MRSA each month. This is roughly twice the number of susceptible individuals colonized with HA-MRSA by the other two groups of patients. However, we examined the WAIC scores of the full model and fourteen sub-models as they fit the data using various priors. We discovered that all these models work nearly equally well, although adding more parameters to the model may boost the WAIC value slightly, but it is difficult to establish which group of patients contributes the most to the MRSA transmission in hospitals. 

Additionally, we simulated the number of new patients colonized with HA-MRSA and infected with HA-MRSA each month from September 2018 to September 2023 under the posterior distribution and compared the results to real-world data from hospitals in Edmonton. The full model and the subset model fit roughly well for the majority of the study period, except for the large spike in the number of new patients colonized with HA-MRSA from July 2023 to September 2023, which demonstrates a weakness in these models that may fail to capture some variations in real-life settings.

The strengths of our study are that we use a simple compartmental model to capture the fundamental epidemiological behaviour of MRSA transmission dynamics in hospitals. It is assumed that individuals in the population are initially susceptible to MRSA, that over time, some of them become colonized, then infected, and that they will eventually be moved to isolated rooms. Using simpler models to describe the complexity of MRSA transmission may help healthcare professionals' understanding and acceptance of these models \citep{mathmodel}. Meanwhile, our model can be computed easily because the mathematical functions are straightforward, illustrated by incorporating priors to estimate parameters in the log-likelihood function. 

Our study shows that compartmental models are statistically helpful in examining MRSA transmission in hospitals \citep{datadriven, stochaMRSA}. It gives us a clear picture of the path that MRSA takes to spread among susceptible people and patients who have been colonized or infected with the bacteria. Both hospitalized HA-MRSA and CA-MRSA patients are included in the model, and their effects on the number of new patients infected with HA-MRSA from susceptible persons are examined. The model results show that all patient categories can raise the risk of MRSA transmission, even if it is challenging to identify which patient group has a greater influence on MRSA transmission.  When we compare the transmission rate of MRSA in Edmonton hospitals to the results in hospitals in the United States using different compartmental models \citep{uspopulationMRSA}, both transmission rates are consistently low. Therefore, other researchers could investigate how MRSA spreads in hospitals in various locations using a model framework similar to this one. Infection control professionals could think about methods like routinely decolonizing patients who are already colonized with HA-MRSA in hospitals, screening patients who are admitted to hospitals with CA-MRSA, putting patients infected with either HA-MRSA or CA-MRSA in isolated rooms, and encouraging everyone in hospitals to wash their hands frequently in order to reduce the risk of MRSA transmission in hospitals.

Some limitations of the study include oversimplifying MRSA transmission processes due to a small number of compartments in the model and failing to account for other factors such as contaminated healthcare workers and environmental impact, despite previous research indicating that these factors influence transmission \citep{nosoMRSA}. The model assumes that the population is homogeneous, implying that each patient has an equal chance of interacting, which does not reflect interactions between patients in different hospitals \citep{betweenhos}. This is because we do not have data for individual hospitals, but rather aggregated data for hospitals in Edmonton, which ignores within-hospital dynamics. We neglected to consider hospital overcrowding, especially during the respiratory season, causing patients to be crammed into hallways since there are not enough beds available, which may raise the risk of MRSA transmission. Nevertheless, there is no seasonal variation in the data over the seasons. In the meantime, we only looked at data from Edmonton, so this model may not accurately reflect what is going on in other parts of Alberta. When we process data, we combine healthcare-associated MRSA cases with community-acquired MRSA cases because they are all identified within the first two days of admission. It may overestimate the impact of patients colonized or infected with CA-MRSA in MRSA transmission in hospitals, because patients could have been colonized or infected with healthcare-associated MRSA in another facility under surveillance. 

Following the approaches used in this study, we may look into MRSA transmission in other parts of Alberta, such as Calgary, the central, north, and south regions, and compare the results to Edmonton. The model might potentially be applied to data from particular hospitals. Fitting models to patient data from other locations or individual hospitals may result in different study conclusions. Using this model, we could then evaluate interventions like screening, decolonization, and hand hygiene for hospitalized MRSA-colonized patients, as well as reducing their contact rate with susceptible people, and measure the change in the overall number of MRSA colonizations or infections over the course of the study. Additionally, since MRSA transmission rates may vary among pediatric, adult, and geriatric populations \citep{riskfactor}, we could categorize patients colonized or infected with HA-MRSA or CA-MRSA into high-risk and low-risk groups based on a number of risk factors, including age and sex. Finally, we could also expand the model structure by taking into account the role that healthcare workers and contamination from environmental bacteria play in hospital MRSA transmission \citep{nosoMRSA}. 
  
  \section*{Data sharing}
   The data that support the findings of this study are available on request from the corresponding author. The data are not publicly available due to privacy or ethical restrictions.

  \section*{Acknowledgements}
   This work was supported by funding from the Natural Sciences and Engineering Research Council of Canada (NSERC) Emerging Infectious Diseases Modelling Initiative (EIDM), “Canadian Network for Modelling Infectious Disease (CANMOD)”, and an Alberta Graduate Excellence Scholarship.

\bibliography{Reference}

\end{document}